\documentclass{revtex4}
\usepackage{amssymb}
\usepackage{amsfonts}
\usepackage{amsmath}

\begin{document}

\title{A new deformed Schi\"{o}berg-type potential and ro-vibrational
energies for some diatomic molecules}
\author{Omar Mustafa}
\email{omar.mustafa@emu.edu.tr}
\affiliation{Department of Physics, Eastern Mediterranean University, G. Magusa, north
Cyprus, Mersin 10 - Turkey,\\
Tel.: +90 392 6301078; fax: +90 3692 365 1604.}

\begin{abstract}
We suggest a new deformed Schi\"{o}berg-type potential for diatomic
molecules. We show that it is equivalent to Tietz-Hua oscillator potential.
We discuss how to relate our deformed Schi\"{o}berg potential to Morse, to
Deng-Fan , to the improved Manning-Rosen, and to the deformed modified
Rosen-Morse potential models. We transform our potential into a proper form
and use the supersymmetric quantization to find a closed form analytical
solution for the ro-vibrational energy levels\ that are highly accurate over
a wide range of vibrational and rotational quantum numbers. We discuss our
results using 4-diatomic molecules NO$\left( X^{2}\Pi _{r}\right) $, O$%
_{2}\left( X^{3}\Sigma _{g}^{-}\right) $, O$_{2}^{+}\left( X^{2}\Pi
_{g}\right) $, and N$_{2}\left( X^{1}\Sigma _{g}^{+}\right) $. Our results
turn out to compare excellently with those from a generalized pseudospectral
numerical method.

\end{abstract}

\maketitle

\section{Introduction}

An empirical diatomic potential energy function provides a quantitative
description of the energy-distance relation that encodes within the relevant
information about a diatomic molecule. Consequently, a large number of
empirical potential models has been suggested \cite{1,2,3,4,5,6,7,8,9,10}.
Improved, extended and/or deformed forms of these potentials were
investigated in the literature \cite%
{11,12,13,14,15,16,17,18,19,20,21,22,23,24,25,26,27,28}. For example, an
extended Lennard-Jones potential is tested by Hajigeorgiou \cite{13}, a
deformed and shifted-by-a-constant Rosen-Morse potential \cite{13,17} is
studied and found to be equivalent to the known Tietz \cite{7} and Wei \cite%
{10} potentials, Wang et al. \cite{12} have shown the equivalence of three
potential models (Manning-Rosen \cite{3}, Schi\"{o}berg \cite{9}, and
Deng-Fan \cite{5}), an improved Schi\"{o}berg potential energy model is
studied by Wang et al. \cite{15}, etc. Nevertheless, the dependence of the
transition probabilities on the rotational-vibrational (ro-vibrational
hereinafter) energy levels has inspired the search for closed-form
analytical energy expressions that are accurate over a broad range of
rotational and vibrational quantum numbers. Such closed-form expressions
provide a substantial simplification of the derivations of the transition
probabilities and are of great advantage in the studies of molecular
transitions in gases, where different collision systems would identify the
gas properties \cite{19}.

However, the main challenge in finding the ro-vibrational energy levels lies
in dealing with the central rotational core, $J\left( J+1\right) /2\mu r^{2}$
with $J\neq 0$, of the radial spherically symmetric Schr\"{o}dinger equation%
\begin{equation}
-\frac{\hbar ^{2}}{2\mu }\frac{d^{2}R_{\nu ,J}\left( r\right) }{dr^{2}}+%
\left[ \frac{J\left( J+1\right) \hbar ^{2}}{2\mu r^{2}}+U\left( r\right) %
\right] R_{\nu ,J}\left( r\right) =E_{\nu ,J}R_{\nu ,J}\left( r\right) ,
\end{equation}%
where $\nu $ denotes the vibrational and $J$ denotes the rotational quantum
numbers. This equation is known to be exactly solvable in a closed form for $%
J=0$. Whereas, for $J\neq 0$ one needs to use an approximation for the
central rotational core term and obtain a closed form analytical solution
(cf, e.g., \cite{22,23}). Hereby, using the Deng-Fan potential \cite{5}%
\begin{equation}
U\left( r\right) =D_{e}\left[ 1-\frac{e^{\lambda r_{e}}-1}{e^{\lambda r}-1}%
\right] ^{2}=A_{1}+\frac{A_{2}}{e^{\alpha r}-1}+\frac{A_{3}}{\left(
e^{\alpha r}-1\right) ^{2}},
\end{equation}%
where%
\begin{equation*}
A_{1}=D_{e};\ A_{2}=-2D_{e}\left( e^{\alpha r_{e}}-1\right) ;\
A_{3}=D_{e}\left( e^{\alpha r_{e}}-1\right) ^{2},
\end{equation*}%
Mustafa \cite{20} has very recently shown (through a quantitative brut-force
numerical test) that the factorization recipe%
\begin{equation}
\frac{r_{e}^{2}}{r^{2}}=C_{0}+\frac{C_{1}}{e^{\lambda r}-1}+\frac{C_{2}}{%
\left( e^{\lambda r}-1\right) ^{2}},
\end{equation}%
of Badawi et al. \cite{21}, for the central rotational core $J\left(
J+1\right) /2\mu r^{2}$, is a more reliable approximation than that of the
improved Greene-Aldrich approximation \cite{22,23}%
\begin{equation}
\frac{1}{r^{2}}\approx \lambda ^{2}\left( \frac{1}{12}+\frac{e^{\lambda r}}{%
\left( e^{\lambda r}-1\right) ^{2}}\right) .
\end{equation}%
It has been observed that the larger the rotational quantum number $J$, the
larger are the ro-vibrational energy shifts/deviations, from the numerically
predicted ones, for a given vibrational quantum number $\nu $ ( for more
details on this issue see Mustafa \cite{20}). In short, the factorization
recipe of Badawi et al. \cite{21} is based on writing the potential and the
central rotational term in homogeneous forms and then determine the
coefficients $C_{i}$'s in terms of the potential parameters.

In the current methodical proposal, we shall use the factorization recipe of
Badawi et al.'s \cite{21} and focus on the derivation of an analytical
expression for the diatomic molecular ro-vibrational energy levels, that has
a sufficiently high accuracy over a relatively broad range of rotational and
vibrational quantum numbers. In so doing, we propose a new (to the best of
our knowledge, of course) 4-parametric deformed Schi\"{o}berg-type \cite%
{9,12,15,16} potential%
\begin{equation}
U\left( r\right) =A\left( B+\tanh _{q}\alpha r\right) ^{2}.
\end{equation}%
Where the $q$-deformation of the usual hyperbolic functions is defined
through%
\begin{equation}
\tanh _{q}x=\frac{\sinh _{q}x}{\cosh _{q}x}\text{; \ }\sinh _{q}x=\frac{%
e^{x}-qe^{-x}}{2}\text{ , \ }\cosh _{q}x=\frac{e^{x}+qe^{-x}}{2}.
\end{equation}%
Here, $A>0,B,q,$ and the screening parameter $\alpha >0$ (which is related
to the potential range) are real\ adjustable parameters to be determined.
Nevertheless, a diatomic molecular potential, necessarily and desirably,
should satisfy the conditions (usually called Varshni's \cite{6} conditions)%
\begin{equation}
\left. \frac{dU\left( r\right) }{dr}\right\vert _{r=r_{e}}=0,\text{ }U\left(
\infty \right) -U\left( r_{e}\right) =D_{e}\text{ },\text{ and \ }\left. 
\frac{d^{2}U\left( r\right) }{dr^{2}}\right\vert _{r=r_{e}}=K_{e}=\left(
2\pi c\right) ^{2}\mu \omega _{e}^{2}.
\end{equation}%
Where $D_{e}$ is the dissociation energy, $r_{e}$ is the equilibrium bond
length, $c$ is the speed of light, $\mu $ is the reduced mass, and $\omega
_{e}$ is the equilibrium harmonic oscillator vibrational frequency.
Moreover, the introduction of a fourth condition $U\left( r_{e}\right) =0$
would only introduce a constant shift (up or down) of the potential curve at
the equilibrium bond length, but never violates the three conditions above 
\cite{11,12,14,15,16,17}. Therefore, the satisfaction of Varshni's
conditions would determine the adjustable parameters $A,B,$ and $q$ of our
new deformed Schi\"{o}berg potential (5).

On the other hand, the Tietz-Hua oscillator potential \cite{18,19}%
\begin{equation}
U\left( r\right) =D_{e}\left[ \frac{1-e^{-b_{h}(r-r_{e})}}{%
1-c_{h}e^{-b_{h}(r-r_{e})}}\right] ^{2}\text{ };\text{ }b_{h}=\beta \left(
1-c_{h}\right) ,\text{ }\beta =\sqrt{\frac{K_{e}}{2D_{e}}},
\end{equation}
is known to be one of the very best analytical potentials in the description
of molecular dynamics at moderate and high vibrational and rotational
quantum numbers. Here, $c_{h}$ represents an optimization parameter obtained
from \emph{ab initio} or Rydberg-Klein-Rees (RKR)\ intermolecular potentials
and $\beta $ is known as the Morse constant (cf., e.g., \cite{18}).
Obviously, in the limit of the optimization parameter $c_{h}\rightarrow 0$,
the Tietz-Hua oscillator potential reduces to the well known Morse
oscillator potential \cite{1}. Moreover, when $c_{h}=e^{-b_{h}r_{e}}$ it
reduces to Deng-Fan potential (2) (which is shown to be equivalent to the
improved Manning-Rosen potential \cite{3,12,27}) with $\lambda =b_{h}$. It
is also a straightforward manner to show that the Tietz-Hua oscillator
potential is equivalent to the deformed modified Rosen-Morse potential \cite%
{14,17,24,25}.

The organization of current work is in order. In section 2, we use Varshni's
conditions (7) to show that our new deformed Schi\"{o}berg-type potential
(5) and the Tietz-Hua oscillator potential (8) are in fact equivalent. In
section 3, we transform our potential (5) into a proper form to be able to
use/recycle the supersymmetric quantization recipe of \cite{24} and obtain a
closed form solution for the ro-vibrational energy levels. We shall,
therefore, only cast the necessary formulae to make the current work
self-contained. We discuss our results, in section 4, using 4-diatomic
molecules NO$\left( X^{2}\Pi _{r}\right) $, O$_{2}\left( X^{3}\Sigma
_{g}^{-}\right) $, O$_{2}^{+}\left( X^{2}\Pi _{g}\right) $, and N$_{2}\left(
X^{1}\Sigma _{g}^{+}\right) $. We compare our results with those of Roy \cite%
{18}, who have used a generalized pseudospectral numerical method (GPS),
whenever possible. We give our concluding remarks in section 5.

\section{Equivalence of the new deformed Schi\"{o}berg potential and the
Tietz-Hua oscillator}

In this section, we shall use Varshni's \cite{6} conditions (7) and show
that our new deformed Schi\"{o}berg-type potential (5) is equivalent to the
Tietz-Hua oscillator potential (8). We start with the application of the
first two conditions, in (7), on our potential (5), i. e., 
\begin{equation}
\left. \frac{dU\left( r\right) }{dr}\right\vert _{r=r_{e}}=0\Longrightarrow
B=-\left( \frac{e^{2\alpha r_{e}}-q}{e^{2\alpha r_{e}}+q}\right) ,
\end{equation}%
and%
\begin{equation}
U\left( \infty \right) -U\left( r_{e}\right) =D_{e}\text{ }\Longrightarrow A=%
\frac{D_{e}}{4q^{2}}\left( e^{2\alpha r_{e}}+q\right) ^{2}.
\end{equation}%
Which when substituted in (5) would yield%
\begin{equation}
U\left( r\right) =D_{e}\left[ 1-\frac{e^{2\alpha r_{e}}+q}{e^{2\alpha r}+q}%
\right] ^{2}.
\end{equation}%
Obviously, the result in (10) implies that the adjustable parameter $A$ is
positive whereas the value of $B$ in (9) can be positive or negative,
depending on the values of the deformation parameter $q$. However, applying
the third condition, \ $\left. d^{2}U\left( r\right) /dr^{2}\right\vert
_{r=r_{e}}=K_{e}$, would imply%
\begin{equation}
\beta =\frac{2\alpha e^{2\alpha r_{e}}}{e^{2\alpha r_{e}}+q}\Longrightarrow
q=-\left( 1-\frac{b}{\beta }\right) e^{br_{e}}\text{ ; \ }b=2\alpha \text{ ,
\ and\ }\beta =\sqrt{\frac{K_{e}}{2D_{e}}}.
\end{equation}%
This would immediately suggest that the deformation parameter $q$ may very
well be defined as%
\begin{equation}
q=-\eta e^{br_{e}}\text{ ; \ }\eta =\left( 1-\frac{b}{\beta }\right) ,
\end{equation}%
where the positivity or negativity of the values of $q$ is determined by the
negativity or positivity of the values of the optimization parameter $\eta $%
, respectively. Moreover, it is clear that the deformation parameter $q$ is
not an $\alpha $-independent but rather a deformation function that depends
on the spectroscopic parameters $b,\beta ,$ and $r_{e}$, i.e., $q\equiv
q\left( b,\beta ,r_{e}\right) \equiv q\left( \alpha ,\beta ,r_{e}\right) $.
This is the only conclusion one can draw form the third condition of (7).

Under such conditional settings, our new deformed Schi\"{o}berg-type
potential (5) (hence equivalently (11)) collapses into the Tietz-Hua
oscillator potential (8)%
\begin{equation}
U\left( r\right) =D_{e}\left[ \frac{1-e^{-b(r-r_{e})}}{1-\eta e^{-b(r-r_{e})}%
}\right] ^{2},
\end{equation}%
where our $\eta =c_{h}$ and $b=2\alpha =b_{h}$ of Kunc et al.'s \cite{19}
and Roy's \cite{18}. The equivalence of our new deformed Schi\"{o}berg-type
potential (5) and the Tietz-Hua oscillator potential (8) is established,
therefore. Yet, the very result of our $q$-deformation in (13) only
documents consistency with the known Tietz-Hua oscillator potential \cite%
{18,19}, where the spectroscopic parameters of the diatomic molecules
(listed in Table 1 below) studied by Kunc et al. \cite{19} and used by Roy 
\cite{18} are readily known and shall be used here as well.

\section{Supersymmetric quantization and ro-vibrational energy levels}

Let us rewrite our new 4-parametric deformed Schi\"{o}berg-type potential
(5) (which is in fact equivalent to that in (11)) as%
\begin{equation}
U\left( r\right) =P_{1}+\frac{P_{2}}{e^{br}+q}+\frac{P_{3}}{\left(
e^{br}+q\right) ^{2}}\text{ ; }b=2\alpha ,
\end{equation}%
where%
\begin{equation}
P_{1}=D_{e}\text{ ; \ }P_{2}=-2D_{e}\left( e^{br_{e}}+q\right) \text{ ; \ }%
P_{3}=D_{e}\left( e^{br_{e}}+q\right) ^{2}.\text{ }
\end{equation}%
At this point, one should notice that Badawi et al. \cite{21} have shown
that the Morse-Pekeris, Rosen-Morse, Manning-Rosen, and Tietz potential
functions are particular cases of the general expression (15). Obviously,
moreover, this potential form (15) represents the first three terms of
Eq.(15) in Jia et al.'s \cite{24} work on a 6-parametric exponential-type
one-dimensional potential. Where we take $P_{4}=P_{5}=0$ of Eq.(15) in \cite%
{24} and interchange the places of $P_{2}$ and $P_{3}$. Hence, the
parametric mapping between Jia et al.'s \cite{24} work and our current
methodical proposal is made clear. As such, we recycle the supersymmetric
quantization recipe of \cite{24} and cast only the necessary formulae to
make the current work self-contained.

Next, we incorporate (15) and (16) into (1) and write the effective
potential as%
\begin{equation}
U_{eff}\left( r\right) =\frac{J\left( J+1\right) \hbar ^{2}}{2\mu r^{2}}%
+U\left( r\right) =\tilde{P}_{1}+\frac{\tilde{P}_{2}}{e^{br}+q}+\frac{\tilde{%
P}_{3}}{\left( e^{br}+q\right) ^{2}},
\end{equation}%
with%
\begin{equation}
\tilde{P}_{1}=P_{1}+\gamma C_{1}\text{ ; \ }\tilde{P}_{2}=P_{2}+\gamma C_{2}%
\text{ };\text{ \ }\tilde{P}_{3}=P_{3}+\gamma C_{3};\text{ }\gamma =\text{\ }%
\frac{J\left( J+1\right) \hbar ^{2}}{2\mu r_{e}^{2}}.
\end{equation}%
Whilst the values of $P_{i}$'s are given in (16), the values of $C_{i}$'s
are obtained using the factorization recipe of Badawi et al. \cite{21} in
the following manner. Let $y=b\left( r-r_{e}\right) $, then with $br=y+u$
and $\,u=br_{e}$ one implies that 
\begin{equation}
\frac{r_{e}^{2}}{r^{2}}=\frac{1}{\left( y/u+1\right) ^{2}}\text{ \ and \ }%
\frac{r_{e}^{2}}{r^{2}}=C_{1}+\frac{C_{2}}{e^{y+u}+q}+\frac{C_{3}}{\left(
e^{y+u}+q\right) ^{2}}\text{.}
\end{equation}%
Retaining the first three terms of the Taylor's expansion near the
equilibrium internuclear distance $y\rightarrow 0$ (i.e., $r\rightarrow
r_{e} $) of both expressions in (19) and equating coefficients of same power
of $y$, one obtains (with $q=-\eta e^{br_{e}}=-\eta e^{u}$) 
\begin{eqnarray}
C_{1} &=&1-\left( \frac{1-\eta }{u}\right) ^{2}\left[ \frac{4u}{1-\eta }%
-\left( 3+u\right) \right] \medskip , \\
C_{2} &=&2e^{u}\left( 1-\eta \right) \left[ 3\left( \frac{1-\eta }{u}\right)
-\left( 3+u\right) \left( \frac{1-\eta }{u}\right) ^{2}\right] \medskip , \\
C_{3} &=&\frac{e^{2u}}{u^{2}}\left( 1-\eta \right) ^{4}\left[ \left(
3+u\right) -\frac{2u}{1-\eta }\right] \medskip .
\end{eqnarray}

Under such potential parametric settings, we may now use the supersymmetric
quantum recipe used by Jia et al.\cite{24} and follow, step-by-step, their
procedure for our Schr\"{o}dinger equation in (5), along with the effective
potential in (17). Namely, one should set their $P_{4}=P_{5}=0$ and their $%
P_{1}$, $P_{3}$, and $P_{2}$ are our current $\tilde{P}_{1}$, $\tilde{P}_{2}$%
, and $\tilde{P}_{3}$, respectively. Hereby, we only cast the necessary
formulae where our superpotential would read%
\begin{equation}
\tilde{W}\left( r\right) =-\frac{\hbar }{\sqrt{2\mu }}\left( \tilde{Q}_{1}+%
\frac{\tilde{Q}_{2}}{e^{br}+q}\right) ,
\end{equation}%
and the one-dimensional ground-state like wave function is given by%
\begin{equation}
\psi \left( r\right) =N\exp \left( -\frac{\sqrt{2\mu }}{\hbar }\int \tilde{W}%
\left( r\right) dr\right) .
\end{equation}%
Which, when substituted in (1) along with (17), would result in%
\begin{equation}
\tilde{Q}_{2}^{2}+bq\,\tilde{Q}_{2}=\frac{2\mu }{\hbar ^{2}}\tilde{P}%
_{3}\Longrightarrow \tilde{Q}_{2}=-\frac{bq}{2}\pm \sqrt{\left( \frac{bq}{2}%
\right) ^{2}+\frac{2\mu }{\hbar ^{2}}\tilde{P}_{3}}
\end{equation}%
\begin{equation}
2\tilde{Q}_{1}\tilde{Q}_{2}-b\tilde{Q}_{2}=\frac{2\mu }{\hbar ^{2}}\tilde{P}%
_{2}\Longrightarrow \tilde{Q}_{1}=\alpha +\frac{\mu \tilde{P}_{2}}{\hbar ^{2}%
\tilde{Q}_{2}}=\frac{1}{2q\tilde{Q}_{2}}\left[ \frac{2\mu }{\hbar ^{2}}%
\left( q\tilde{P}_{2}+\tilde{P}_{3}\right) -\tilde{Q}_{2}^{2}\right] ,
\end{equation}%
and%
\begin{equation}
\tilde{Q}_{1}^{2}=\frac{2\mu }{\hbar ^{2}}\left( \tilde{P}_{1}-E_{0}\right)
\Longrightarrow E_{0}=\tilde{P}_{1}-\frac{\hbar ^{2}}{2\mu }\left( \frac{1}{%
2q\tilde{Q}_{2}}\left[ \frac{2\mu }{\hbar ^{2}}\left( q\tilde{P}_{2}+\tilde{P%
}_{3}\right) -\tilde{Q}_{2}^{2}\right] \right) ^{2}.
\end{equation}%
Hence, the wave function reads%
\begin{equation}
\psi \left( r\right) =N\,e^{\tilde{Q}_{1}r}\left( \frac{e^{br}}{e^{br}+q}%
\right) ^{\tilde{Q}_{2}/bq}
\end{equation}%
and the corresponding ro-vibrational energy levels (with $b=2\alpha $) are%
\begin{equation}
E_{\nu ,J}=\tilde{P}_{1}-\frac{\hbar ^{2}b^{2}}{2\mu }\left[ \frac{\frac{%
2\mu }{\hbar ^{2}q^{2}b^{2}}\left( \tilde{P}_{3}+q\tilde{P}_{2}\right) }{%
-1-2\nu \pm \sqrt{1+\frac{8\mu }{\hbar ^{2}q^{2}b^{2}}\tilde{P}_{3}}}-\frac{%
\left( -1-2\nu \pm \sqrt{1+\frac{8\mu }{\hbar ^{2}q^{2}b^{2}}\tilde{P}_{3}}%
\right) }{4}\right] ^{2},
\end{equation}%
The positive and negative sings $\left( \pm \right) $, however, correspond
to positive and negative values of $q$, respectively. That is, one takes the
positive sing for $q>0$ and the negative sign for $q<0$. This would, in
turn, ensure that the wavefunction (28) vanishes as $r\rightarrow \infty $
and becomes finite as $r\rightarrow 0$. For more details on this issue the
reader may refer to Jia et al.\cite{24}.

\section{Results and Discussion}

The spectroscopic parameters for 4-diatomic molecules NO$\left( X^{2}\Pi
_{r}\right) $, O$_{2}\left( X^{3}\Sigma _{g}^{-}\right) $,\ O$_{2}^{+}\left(
X^{2}\Pi _{g}\right) $ and N$_{2}\left( X^{1}\Sigma _{g}^{+}\right) $ as
reported by Kunc et al. \cite{19} are summarized in table 1. We now use our
result in (29) and calculate the ro-vibrational energy levels listed in
table 2 for NO$\left( X^{2}\Pi _{r}\right) $, O$_{2}\left( X^{3}\Sigma
_{g}^{-}\right) $, O$_{2}^{+}\left( X^{2}\Pi _{g}\right) $ molecules, and
the vibrational energies for the N$_{2}\left( X^{1}\Sigma _{g}^{+}\right) $
molecule in table 3. For each of the given diatomic molecules above we have
tested the sign of $q$ and accordingly used the proper sign of the square
root in (29).

In table 2, we compare our results with those of Roy \cite{18}, who have
used a GPS numerical method. Roy's results compared excellently with the
results of the Nikiforov-Uvarov formalism of Hamzavi et al. \cite{28}.
Moreover, in the conversion of the $\left( eV\right) $-units used by Roy 
\cite{18} into $\left( cm^{-1}\right) $-units, we have used the relation%
\begin{equation*}
E_{\nu ,J}\left( cm^{-1}\right) =D_{e}\left( cm^{-1}\right) +\frac{%
Roy^{\prime }s\left( eV\right) }{1.23941188\times 10^{-4}\left(
eV/cm^{-1}\right) }.
\end{equation*}%
It is obvious that our results obtained form (29) are in excellent agreement
with those from the GPS numerical method \cite{18} (hence, with the results
of the Nikiforov-Uvarov formalism of Hamzavi et al. \cite{28}), whenever
available, of course. Yet, in the search for any connection between the
accuracy of our results reported in table 2 and the potential parameters
listed in table 1, we observe a general trend that the heavier/larger the
reduced mass the more accurate our result are compared to GPS ones. This is
very much related to the semiclassical limit nature (similar recipes were
used early on, like the known large-$\ell $ expansion technique, cf. e.g., 
\cite{29,30,31}) of the Taylor's expansion near the equilibrium internuclear
distance, $r\rightarrow r_{e}$, used in the factorization recipe (19) of
Badawi et al. \cite{21}. It is also obvious that the larger the reduced
mass, in the central core term $J\left( J+1\right) /2\mu r^{2}$, the less
the effect of the rotational quantum number $J$. The factorization recipe
(19) of Badawi et al. \cite{21} is indeed an excellent approximation for the
ro-vibrational energy levels.

Furthermore, the authors of \cite{14,17} have used the common potential (11)
(i.e., their Eq.(11) in both \cite{14,17}) as an equivalent form for their
deformed modified Rosen-Morse potential. Therefore, the introduction of
table 3 is unavoidably in the process. In this table we compare our results
with those reported by Lino da Silva et al. \ \cite{26} (who have used the
RKR method to construct the potential curve of the N$_{2}\left( X^{1}\Sigma
_{g}^{+}\right) $) along with the results reported by Sun et al. \cite{17}
and those of Morse potential (i.e., for $\eta =c_{h}\rightarrow 0$). The
comparison between our results and those of Lino da Silva et al.\ \cite{26}
shows that the accuracy is still high (i.e., the accuracy is $\sim 99.1\%$
and is better than that from the Morse, especially for large vibrational
quantum numbers $\nu $). However, when we compare our results with those of
Sun et al. \cite{17}, we observe small discrepancies. These discrepancies
are very much related to the improper mathematical argument used by Sun et
al. \cite{17} (more details on this issue are discussed in the Appendix
below). Moreover, it is a straightforward manner to show that our energy
expression (29) is in exact accord with that of Eq.(14) of Sun et al. \cite%
{17} for the case $J=0$ they have considered.

\section{Concluding remarks}

In this work, we have introduced a new (to the best of our knowledge)
deformed Schi\"{o}berg-type potential (5). We have shown that upon the
application of Varshni's \cite{6} conditions (7), our deformed potential (5)
collapses into a general/common form (11) shared by a number of well known
diatomic potential models. For example, for \ the Tietz-Hua oscillator
potential $q=-\eta e^{br_{e}}$, $\eta =c_{h}$, and $2\alpha =b=b_{h}$ \cite%
{18,19}, for the Morse oscillator potential \cite{1} $\eta \rightarrow 0$
(i.e., the deformation parameter $q\rightarrow 0$), for the Deng-Fan
potential (2) and the improved Manning-Rosen potential \cite{3,12,27} $q=-1$
(i.e., $\eta =e^{-br_{e}}$), etc.

To find a highly accurate (over a wide range of vibrational and rotational
quantum numbers) analytical expression for the ro-vibrational energy levels,
we have adopted/favoured Badawi et al.'s \cite{21} factorization recipe (to
deal with the central rotational core $J\left( J+1\right) /2\mu r^{2}$) and
recycled the supersymmetric quantization approach used by Jia et al. \cite%
{24} for a 6-parametric exponential potential model. Our strategy was
inspired by Badawi et al.'s \cite{21} work on writing the potential and the
central rotational term in homogeneous forms (i.e., (17) and (19)) and then
determine the coefficients $C_{i}$'s in terms of the potential parameters to
workout an analytical expression (29) for the ro-vibrational energy levels.
For the 3-diatomic molecules NO$\left( X^{2}\Pi _{r}\right) $, O$_{2}\left(
X^{3}\Sigma _{g}^{-}\right) $, and O$_{2}^{+}\left( X^{2}\Pi _{g}\right) $
we have used, our results turned out to be highly accurate compared with the
numerically predicted ones of Roy \cite{18}, who have used a generalized
pseudospectral method (GPS) (documented in table 2).

\section{Appendix: Improper determination of the spectroscopic parameter $%
\protect\alpha $}

For the sake of mathematical correctness and scientific honesty, it is
deemed unavoidable to introduce table 3 and to pinpoint the mathematical
mismanagement in the determination of the spectroscopic parameter $\alpha $
committed by the authors of \cite{14,17}.

In their attempt to determine the spectroscopic parameter $\alpha $, the
authors of \cite{14,17} have, mathematical wise, mishandled equation (12).
They have considered that the deformation parameter $q$ is $\alpha $%
-independent and used $\alpha $ values in table 1 to obtain $q$ in (12) (as
documented in their $q$ values listed in their table 1 of \cite{14} and used
again in \cite{17}). That is, they admit that $q=-\eta
e^{br_{e}}=-c_{h}e^{br_{e}};$ $\alpha =b/2$ as given by (12) but again they
have used the very same values of $\alpha $ in table 1 to find $\alpha
=\alpha _{DMRM}$ (below). This is an odd and/or improper mathematical
treatment. Yet, their result (as documented in their Eq.(23) of \cite{17}
and Eq.(26) of \cite{14}) 
\begin{equation*}
\alpha _{DMRM}=\frac{1}{2}\beta +\frac{1}{2r_{e}}W\left( r_{e}q\beta
e^{-r_{e}\beta /2}\right) \text{ ; \ }\beta =\sqrt{\frac{K_{e}}{2D_{e}}},
\end{equation*}%
should be corrected into%
\begin{equation*}
\alpha _{DMRM}=\frac{1}{2}\beta +\frac{1}{2r_{e}}W\left( r_{e}q\beta
e^{-r_{e}\beta }\right) \text{,}
\end{equation*}%
\ where $W$ is the Lambert function. However, their approach would remain
improper. 

\newpage

\begin{center}
\bigskip 
\begin{tabular}{l}
Table 1: \\ \hline
\multicolumn{1}{c}{%
\begin{tabular}{cccccccc}
Molecule & $\eta $ & $\mu /10^{-23}\left( g\right) $ & $\alpha \left( 
\mathring{A}^{-1}\right) $ & $r_{e}\left( \mathring{A}\right) $ & $\beta
\left( \mathring{A}^{-1}\right) $ & $D_{e}\left( cm^{-1}\right) $ & $\omega
_{e}\left( cm^{-1}\right) $ \\ \hline
NO$\left( X^{2}\Pi _{r}\right) $ & -0.029477 & 1.249 & 1.357795 & 1.151 & 
2.7534 & 53341 & 1904.2 \\ 
O$_{2}\left( X^{3}\Sigma _{g}^{-}\right) $ & 0.027262 & 1.377 & 1.295515 & 
1.207 & 2.6636 & 42041 & 1580.2 \\ 
O$_{2}^{+}\left( X^{2}\Pi _{g}\right) $ & -0.019445 & 1.377 & 1.434935 & 
1.116 & 2.8151 & 54688 & 1904.8 \\ 
N$_{2}\left( X^{1}\Sigma _{g}^{+}\right) $ & -0.032325 & 1.171 & 1.392925 & 
1.097 & 2.6986 & 79885 & 2358.6 \\ \hline
\end{tabular}%
}%
\end{tabular}%
\bigskip

\begin{tabular}{l}
Table 2: \\ \hline
\multicolumn{1}{c}{%
\begin{tabular}{cccccccc}
$\ $ &  & NO$\left( X^{2}\Pi _{r}\right) $ & $\ $ & O$_{2}\left( X^{3}\Sigma
_{g}^{-}\right) $ & \ \ \  & O$_{2}^{+}\left( X^{2}\Pi _{g}\right) $ &  \\ 
\hline
$\nu $\  & $J$ & GPS \cite{18}\ \  & Eq.(29) \ \ \  & GPS \cite{18}\ \  & 
Eq.(29) \ \ \  & GPS \cite{18}\ \  & Eq.(29) \\ \hline
0 & 0 & 947.759 & 947.756 & 774.984 & 775.089 & 934.601 & 934.614 \\ 
& 1 & 951.123 & 951.121 & 777.848 & 777.863 & 937.848 & 937.862 \\ 
& 2 & 957.849 & 957.847 & 783.394 & 783.410 & 944.341 & 944.353 \\ 
& 3 &  & 967.937 &  & 791.731 &  & 954.094 \\ 
& 4 &  & 981.390 &  & 802.823 &  & 967.079 \\ 
& 5 &  & 998.205 &  & 816.688 &  & 983.310 \\ 
& 10 & 1132.686 & 1132.686 & 927.562 & 927.578 & 1113.112 & 1113.127 \\ 
& 15 & 1351.069 & 1351.072 & 1107.634 & 1107.654 & 1323.924 & 1323.940 \\ 
& 20 & 1653.146 & 1653.153 & 1356.714 & 1356.739 & 1615.541 & 1615.563 \\ 
\hline
3 & 0 & \ \ \ 6453.267 \ \  & 6453.239 & \ \ \ 5269.581 \ \  & 5269.672 & \
\ \ 6376.545 \ \  & 6376.615 \\ 
& 1 & 6456.510 & 6456.484 & 5272.250 & 5272.343 & 6379.684 & 6379.756 \\ 
& 2 & 6462.995 & 6462.971 & 5277.588 & 5277.684 & 6385.962 & 6386.035 \\ 
& 3 &  & 6472.703 &  & 5285.694 &  & 6395.455 \\ 
& 4 &  & 6485.677 &  & 5296.374 &  & 6408.015 \\ 
& 5 &  & 6501.894 &  & 5309.722 &  & 6423.713 \\ 
& 10 & 6631.552 & 6631.592 & 5416.325 & 5416.479 & 6549.135 & 6549.270 \\ 
& 15 & 6842.080 & 6842.207 & 5589.607 & 5589.837 & 6752.948 & 6753.159 \\ 
& 20 & 7133.275 & 7133.526 & 5829.279 & 5829.619 & 7034.867 & 7035.194 \\ 
\hline
5 & 0 & \ \ \ 9951.736 \ \  & 9951.693 & \ \ 8118.378 \ \  & 8118.516 & \ \
\ 9845.984 \ \  & 9846.089 \\ 
& 1 & 9954.898 & 9954.857 & 8120.977 & 8121.118 & 9849.051 & 9849.159 \\ 
& 2 & 9961.220 & 9961.188 & 8126.175 & 8126.321 & 9855.183 & 9855.296 \\ 
& 3 &  & 9970.679 &  & 8134.126 &  & 9864.503 \\ 
& 4 &  & 9983.3351 &  & 8144.530 &  & 9876.778 \\ 
& 5 &  & 9999.155 &  & 8157.535 &  & 9892.120 \\ 
& 10 & 10125.542 & 10125.669 & 8261.257 & 8261.546 & 10014.566 & 10014.830
\\ 
& 15 & 10330.775 & 10331.112 & 8429.966 & 8430.441 & 10213.639 & 10214.091
\\ 
& 20 & 10614.632 & 10615.269 & 8663.303 & 8664.046 & 10488.989 & 10489.719
\\ \hline
\end{tabular}%
}%
\end{tabular}%
\bigskip 

\begin{tabular}{l}
Table 3: \\ \hline
\multicolumn{1}{c}{%
\begin{tabular}{ccccc}
\hline
$\nu $\  & RKR \cite{26} & DMRM \cite{17} \ \ \  & Eq.(29)\ \  & Morse \cite%
{17} \ \ \  \\ \hline
0 & 1184.4539 & 1174.9971 & 1174.9270 & 1174.9477 \\ 
1 & 3526.3576 & 3499.8409 & 3499.7430 & 3498.7289 \\ 
2 & 5833.4516 & 5790.8755 & 5790.7601 & 5787.6913 \\ 
3 & 8107.0460 & 8048.0809 & 8047.9316 & 8041.8351 \\ 
4 & 10348.312 & 10271.387 & 10271.210 & 10261.160 \\ 
5 & 12558.287 & 12460.752 & 12460.549 & 12445.666 \\ 
6 & 14737.876 & 14616.138 & 14615.901 & 14595.353 \\ 
7 & 16887.859 & 16737.473 & 16737.218 & 16710.222 \\ 
8 & 19008.895 & 18824.747 & 18824.454 & 18790.272 \\ 
9 & 21101.519 & 20877.869 & 20877.559 & 20835.503 \\ \hline
\end{tabular}%
}%
\end{tabular}

\medskip \medskip
\end{center}

\textbf{Tables captions:}

\textbf{Table 1:} Spectroscopic parameters and constants used in this work,
quoted from \cite{18}.

\textbf{Table 2: }Ro-vibrational energies $E_{\nu ,J}$ (in $cm^{-1}$ units)
for 3-diatomic molecules with $\nu =0,3,5$ and different value of $J$. Our
results in (29) are compared with those of Roy's \cite{18} (GPS) whenever
possible.

\textbf{Table 3: }Vibrational energies $E_{\nu ,J}$ (in $cm^{-1}$ units) for
N$_{2}\left( X^{1}\Sigma _{g}^{+}\right) $. Our results (29) are compared
with those of \ RKR \cite{26}, DMRM \cite{17}, and Morse \cite{17}.


\newpage 

\begin{thebibliography}{99}
\bibitem{1} P M Morse, Phys. Rev. \textbf{34} (1929) 57.

\bibitem{2} N. Rosen, P. M. Morse, Phys. Rev. \textbf{42} (1932) 210.

\bibitem{3} M. F. Manning, N. Rosen, Phys. Rev. \textbf{44} (1933) 951.

\bibitem{4} A. A. Frost, B. Musulin, J. Chem. Phys. \textbf{22} (1954) 1017.

\bibitem{5} Z. H. Deng, Y. P. Fan, Shandong Univ. J. \textbf{7} (1957) 162.

\bibitem{6} Y P Varshni, Rev. Mod. Phys. \textbf{29} (1957) 664. S
Noorizadeh,

G R Pourrshams, J. Mol. Structure (Theochem) \textbf{678 }(2004) 207.

\bibitem{7} T. Tietz, J. Chem. Phys. \textbf{38} (1963) 3036.

\bibitem{8} J. M. Murrell, K. S. Sorbie, J. Chem. Soc. Faraday Trans. 
\textbf{70} (1974) 1552.

\bibitem{9} D. Schi\"{o}berg, Mol. Phys. \textbf{59} (1986) 1123.

\bibitem{10} H. Wei, Phys. Rev. \textbf{A42} (1990) 2524.

\bibitem{11} C-S Jia et al., J. Chem. Phys. \textbf{137} (2012) 014101.

\bibitem{12} P-Q Wang, L-H Zhang, C-S Jia, J-Y Liu, J. Mol. spectrosc.%
\textbf{\ 274} (2012) 5.

\bibitem{13} P. G. Hajigeorgiou, J. Mol. spectrosc.\textbf{\ 263} (2010)
101.{}

\bibitem{14} C-S Jia, T. Chen, L-Z Yi, S-R Lin, J. Math. Chem. \textbf{51}
(2013) 2165.

\bibitem{15} P-Q Wang et al, J. Mol. spectrosc.\textbf{\ 278} (2012) 23.

\bibitem{16} J. Lu, Phys. Scr. \textbf{72 }(2005) 349.

S. Ortakaya, Few Body Syst. \textbf{54} (2013) 1901.

\bibitem{17} Y. Sun, S He, C-S Jia, Phys. Scr. \textbf{87} (2013) 025301.

\bibitem{18} A K Roy, J. Math. Chem. \textbf{52} (2014) 1405.

\bibitem{19} J A Kunc, F J Gordillo-V\'{a}zquez, J. Phys. Chem. \textbf{101}
(1997) 1595.

\bibitem{20} O. Mustafa, J. Phys. \textbf{B}: At. Mol. Opt. Phys. \textbf{48}
(2015) 065101.

\bibitem{21} M Badawi, N Bessis, G Bessis, J. Phys. \textbf{B}: atom. Molec.
Phys. \textbf{5} (1972) L157.

\bibitem{22} R L Greene, C Aldrich, Phys. Rev. \textbf{A 14} (1976) 2363

\bibitem{23} J-Y Liu, G-D Zhang, C-S Jia, Phys. Lett. \textbf{A377} (2013)
1444.

\bibitem{24} C-S Jia et al., Commun. Theor. Phys. \textbf{37} (2002) 523.

\bibitem{25} H. Egrifes, D Demirhan, F. Buyukkilic, Phys. Scr. \textbf{60}
(1999) 195.

\bibitem{26} M. Lino da Silva, V. Guerra, J. Loureiro, P. A. Sa, Chem. Phys. 
\textbf{348} (2008) 187.

\bibitem{27} X-T Hu, L-H Zhang, C-S Jia, J. Mol. spectrosc.\textbf{\ 297}
(2014) 21.

\bibitem{28} M. Hamzavi, A. A. Rajabi, K-E Thylwe, Int. J. Quantum Chem. 
\textbf{112} (2012) 2701.

\bibitem{29} O. Mustafa and M. Odeh, J. Phys. \textbf{A33} (2000) 5207.

\bibitem{30} M. Znojil, F. Gemperle, O. Mustafa; J. Phys. \textbf{A35}
(2002) 5781.

\bibitem{31} O Mustafa and M. Znojil; J. Phys. \textbf{A35} (2002)\textbf{\ }%
8929\textbf{.}
\end{thebibliography}
\end{document}